# First results of a large-area cryogenic gaseous photomultiplier coupled to a dual-phase liquid xenon TPC

L. Arazi[a,*], A. E. C. Coimbra[a,b,†], E. Erdal[a], I. Israelashvili[a,c], M. L. Rappaport[a], S. Shchemelinin[a], D. Vartsky[a], J. M. F. dos Santos[b], and A. Breskin[a]

[a] *Weizmann Institute of Science,*
    *Rehovot 76100, Israel*
[b] *University of Coimbra,*
    *Coimbra, Portugal*
[c] *Nuclear Research Center Negev,*
    *Beer-Sheva 9001, Israel*

    *E-mail*: lior.arazi@weizmann.ac.il

ABSTRACT: We discuss recent advances in the development of cryogenic gaseous photomultipliers (GPM), for possible use in dark matter and other rare-event searches using noble-liquid targets. We present results from a 10 cm diameter GPM coupled to a dual-phase liquid xenon (LXe) TPC, demonstrating − for the first time − the feasibility of recording both primary ("S1") and secondary ("S2") scintillation signals. The detector comprised a triple Thick Gas Electron Multiplier (THGEM) structure with cesium iodide photocathode on the first element; it was shown to operate stably at 180 K with gains above $10^5$, providing high single-photon detection efficiency even in the presence of large alpha particle-induced S2 signals comprising thousands of photoelectrons. S1 scintillation signals were recorded with a time resolution of 1.2 ns (RMS). The energy resolution ($\sigma/E$) for S2 electroluminescence of 5.5 MeV alpha particles was ~9%, which is comparable to that obtained in the XENON100 TPC with PMTs. The results are discussed within the context of potential GPM deployment in future multi-ton noble-liquid detectors.

---

[*] Corresponding author
[†] Equal contributor

**Contents**

# 1. Introduction

Dual-phase noble-liquid time projection chambers (TPCs) [1-3] employing liquid xenon (LXe) or liquid argon (LAr) as their target material are at the forefront of direct searches for dark matter (DM) in the form of weakly interacting massive particles (WIMPs). While running experiments utilize ~30-120 kg of noble liquid as their fiducial targets [4-7], ton-scale experiments are already being constructed [8, 9] or planned for the next few years [10-13]; multi-ton experiments such as DARWIN [14, 15] are expected to take the stage in the 2020s and probe the WIMP parameter space down to the region where coherent neutrino-nucleus scattering becomes the dominant background.

In dual-phase noble-liquid DM detectors, particle interactions within the noble liquid lead to two UV-emission processes: a prompt scintillation signal ("S1") emanating from the interaction site and a delayed electroluminescence signal ("S2"), generated by the ionization electrons liberated in the interaction, as they cross the high field region in the vapor phase after being extracted from the liquid [2]. These two signals are used for estimating the deposited energy, localize the event within the active target, and discriminate between candidate WIMP events and background produced by electron recoils. Presently, all noble-liquid detectors for direct dark matter searches – both single-phase and dual-phase – employ vacuum photomultiplier tubes (PMTs) for the detection of the UV signals. In single-phase liquid-only experiments the entire wall area is covered by PMTs [16, 17]; in dual-phase experiments, the PMTs are arranged in two arrays, one at the top and one at the bottom the TPC, with the walls covered by UV-reflective PTFE [18-20]. The PMTs for dark matter searches are made of low-radioactivity materials, operate in cryogenic conditions, sustain high pressures, and have high quantum efficiency (QE) at the desired wavelength (>30%), and thus provide an adequate solution for the current experimental requirements. However, the challenging goals of future multi-ton experiments motivate the development of new photon-detector concepts; these should not only be more affordable than PMTs, but must also allow for a significant improvement in detection sensitivity and background rejection.

Cryogenic Gaseous Photomultipliers (GPM [21]) may constitute an economic high-performance alternative to PMTs for future large-scale noble-liquid detectors. GPMs are gaseous detectors, in which incoming photons release photoelectrons from a photocathode of high quantum efficiency; these are subsequently focused into a region of a strong electric field, where they undergo avalanche multiplication, enabling high single-photon detection and localization capabilities. GPMs comprising a reflective UV-sensitive cesium iodide (CsI) photocathode [22] coupled to a wire chamber have already been successfully employed in room-temperature experiments for the identification of relativistic particles as Ring Imaging Cherenkov (RICH) devices [23]. More recently, GPMs developed at the Weizmann Institute, relying on a triple-GEM (Gas Electron Multiplier [24]) structure with a reflective CsI coating on the first amplification stage [25], were successfully employed as RICH devices in the Hadron Blind Detector of PHENIX [26-28]. CsI-coated Thick-GEM (THGEM) based GPMs [29-33] are presently under advanced development for the upgrade of the RICH-1 detector of the COMPASS experiment [34, 35]. In addition, a quintuple-GEM RICH detector prototype with a reflective CsI photocathode has recently undergone its first successful beam tests in SLAC and FNAL [36].

In the context of DM detection, cryogenic GPMs may allow for large-area coverage with high detection efficiency and high filling factor (due to the possibility of having large square or hexagonal modules) - potentially at lower cost than PMTs; made of suitable materials, they could reach similar radiopurity levels. The geometry of the multiplier electrodes, choice of gas mixture, operating pressure, module size, readout scheme and pixel size can be tailored to meet specific experimental requirements. In dual-phase TPCs, windowed GPMs deployed in nearly-$4\pi$ coverage (as suggested, for example, in [37]) would potentially increase the detector's sensitivity to low-energy depositions, and hence to low-mass WIMPs.

Previous experiments with a 30 mm diameter double-THGEM GPM and a GPM consisting of a hybrid structure of a THGEM, Parallel Ionization Multiplier (PIM) and MICROMEGAS, demonstrated the feasibility of recording primary scintillation light from the tracks of alpha particles emitted into liquid xenon [38, 39]. In what follows, we describe our recent results in operating a 100 mm diameter, triple-THGEM cryogenic GPM prototype with a reflective CsI photocathode, coupled to a small dual-phase LXe TPC through a fused-silica window. These results represent the first demonstration of the ability of such a detector to record both single photons and massive alpha particle-induced S2 signals comprising thousands of photoelectrons, in the same operating conditions. Experimental results regarding the detector gain and stability, and energy and time resolutions are presented. While the photon detection efficiency of the detector was not measured directly, we provide estimates regarding its value in the present prototype and discuss prospects for its enhancement. We further outline the potential application and advantages of GPMs deployed in a dual-phase $4\pi$-TPC configuration, emphasizing the main challenges which must be overcome for realizing this scheme.

## 2. Experimental setup and procedures

The experiments were conducted using the WILiX LXe cryostat, described in detail in [40]. For the present study, the cryostat (shown schematically in figure 1), was used to house a small dual-phase TPC at its center (figures 1 and 2a). Two electroformed Cu meshes with 85% transparency (Precision Eforming, MC17), set 5 mm apart, were used as the TPC anode and gate electrodes - bounding the liquid-gas interface from above and below, respectively. The liquid level was controlled by a movable "weir" (not shown in the figure). An 18 mm diameter stainless steel disc, serving both as the TPC cathode and alpha particle source substrate, was suspended 5 mm below the gate mesh. The disc carried a central oval (~8×5 mm) active spot of 80 Bq $^{241}$Am (spectral resolution ~4% FWHM). Prior to the experiments, the source was tested in liquid nitrogen, showing no loss of activity in repeated thermal cycles. The TPC voltages, set by CAEN N1471H power supplies, were −3 kV for the cathode/source, −2.5 kV for the gate and +2 kV for the anode, defining a drift field of 1 kV/cm and a nominal extraction field of 12 kV/cm in the gas phase (assuming the liquid-gas interface lies half-way between the meshes). Alpha emissions from the source into the liquid resulted in prompt S1 light signals, followed by secondary S2 signals appearing 2.4 μs later. A 1" square PMT (Hamamatsu R8520-06-Al), located 3.5 cm below the source, was used to record reflected S1 and S2 photons. Throughout the experiments the xenon pressure in the TPC was 1.8 bar absolute and the recirculation flow was 3 slpm.

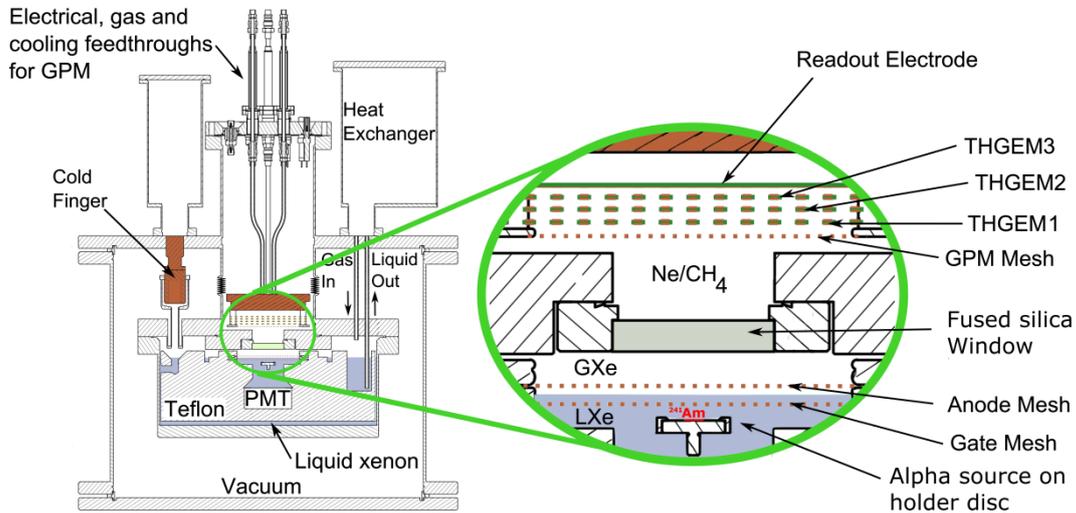

**Figure 1:** Left - the liquid xenon cryostat (WILiX), including the GPM assembly and inner dual-phase TPC; right – enlarged view of the GPM and TPC region.

The GPM assembly was installed in the central port of WILiX, as shown in figure 1, above a UV transparent window with a 35.6 mm aperture (MPF A0650-2-CF, with Corning HPFS 7980 fused silica). The window transmission was measured to be 90% at 175 nm. The GPM prototype investigated in this work, shown in detail in figure 2, comprised a cascaded structure of three THGEM electrodes, with an active diameter of 100 mm. They were made of an FR4 plate with a thickness $t$ = 0.4 mm, Cu-clad on both sides. The drilled hole pattern was hexagonal, with a pitch $a$ = 0.8 mm (between the hole centers) and hole diameter $d$ = 0.4 mm; the width of the etched hole rims ($h$) was 50 μm. The Cu layer thickness (after etching) was 64 μm. The THGEM electrodes were produced by ELTOS SpA, Italy. The final processing stages, including gold-plating, cleaning and baking were done in the CERN MPGD workshop (gold plating was applied to provide a suitable substrate for the CsI photocathode [22]). For the first experiments, described in this work, the three THGEM electrodes were mounted with their holes aligned. The transfer gaps between the stages, as well as the induction gap between THGEM3 and the un-segmented readout anode were 2 mm wide. Each of the THGEM faces had a separate HV bias. An electroformed Cu mesh with 85% transparency (Precision Eforming, MC17) was mounted 3 mm below THGEM1 and kept at the same potential as its CsI-coated face, to maximize the extraction efficiency of electrons from the reflective photocathode [41]. HV bias was provided through low-pass filters, using CAEN N1471H power supplies. Signals were extracted from the anode through a coaxial cable into a Canberra 2006 charge sensitive preamplifier connected on the outer side of the GPM chamber's top flange (which included all other GPM electrical feedthroughs; see figure 1). All internal HV wires (Allectra 311-KAP1) were Kapton-coated and rated for ultra-high vacuum. The vacuum reached in the GPM chamber after its installation and just before introducing the counting gas was ~$2\times10^{-6}$ torr.

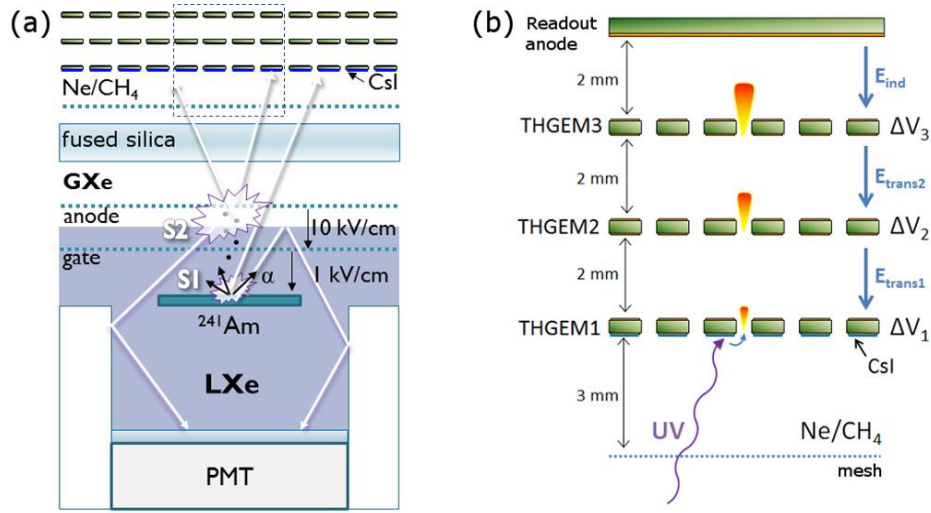

**Figure 2:** (a) Schematic drawing of the experiment, showing alpha particle-induced S1 and S2 signals, detected directly by the GPM (above a fused silica window) and, through reflections, by the bottom PMT. (b) Enlarged view of the dashed region in (a) showing the triple-THGEM GPM with CsI photocathode on its first stage; $\Delta V_k$ is the voltage across the k-th stage. $E_{trans1,2}$ are the nominal transfer fields between the amplification stages and $E_{ind}$ is the induction field. The CsI-coated face of THGEM1 and the mesh are held at the same potential.

In order to control the temperature of the GPM gas, the detector was mounted below a Cu block cooled by liquid nitrogen vapor (figure 3a). The block temperature, measured by a Pt100 sensor, was controlled by a 50 W heater powered by a Cryo-con Model 24C temperature controller. The incoming GPM counting gas (Ne/$CH_4$ mixtures) passes through the Cu block on a "serpentine" path to maximize the heat exchange. Most of the data were taken at a pressure of 0.7 bar and 180 K, corresponding to roughly the same gas density as in 1.1 bar at room temperature. In some cases, the Cu block was not actively cooled by liquid nitrogen; it was found that in this mode the thermal conductivity of the gas maintained the avalanche region of the GPM at ~190 K, with 220 K on the Cu block. No significant changes were observed in the detector performance at 180 K and 190 K.

Ne/$CH_4$ was chosen as the GPM gas because it provides both high gas gain at relatively low voltages [42] and high photoelectron extraction efficiency from the CsI photocathode [31, 43]. For the initial experiments described in this work, we used Ne/$CH_4$(5%), Ne/$CH_4$(10%), and Ne/$CH_4$(20%), operating the detector in a sealed mode (i.e., with no gas flow).

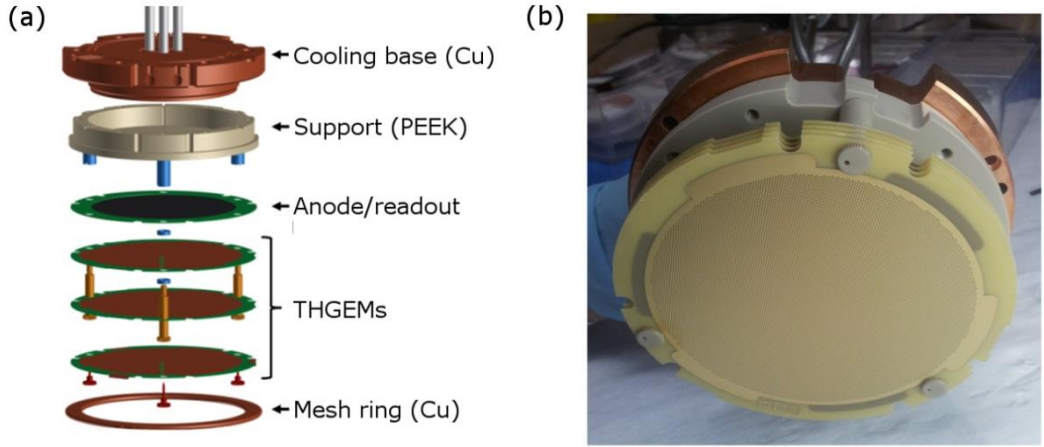

**Figure 3:** (a) The GPM assembly, including the Cu cooling base (see text). (b) A photograph of the GPM prototype before wiring.

CsI was evaporated onto THGEM1 in a dedicated system. The procedure was based on previous experience [22], but without heating the photocathode after evaporation, as it was found that this did not affect its QE or stability against short exposures to air. The deposited layer was ~300 nm-thick; the typical QE value at 175 nm, measured for several photocathodes during the study (using a McPherson 302 monochromator), was ~20-25%. After evaporation, the CsI-coated THGEM was transferred in a closed box into a $N_2$-filled glove box, where it was mounted on the GPM assembly; the complete assembly was subsequently installed under a flow of Ar inside the cryostat (figure 1).

## 3. Results

### 3.1 Observing S1 and S2 signals with the GPM

At adequate electric fields in the LXe TPC, alpha-particle induced S1 and S2 signal pairs were observed on the GPM anode, starting at gains of ~$10^3$. For a drift field of 1 kV/cm in LXe and nominal extraction field of 12 kV/cm in the gas phase, S2 signals appeared ~2.4 μs after S1, as expected based on the known drift velocity of electrons in LXe [44, 45]. The ratio of S2 to S1 pulse areas measured by the PMT was ~25 (the PMT signal was used as direct input to the oscilloscope without amplification and shaping; the pulse area was therefore proportional to the number of photoelectrons). Figure 4 shows a typical signal of the charge sensitive preamplifier (CSP) connected to the GPM anode, along with the corresponding PMT signal for the same event. In this particular case the GPM was operated with Ne/$CH_4$(10%) at a pressure of 1.05 bar at ~190 K. The voltage across THGEM1 was 1250 V, with 1050 V across THGEMs 2 and 3 (overall gain of ~$1\times10^5$); the transfer and induction fields were 0.5 kV/cm. Figure 5 shows an example of the GPM signal where the CSP output was processed by a timing filter amplifier (ORTEC 474), with integration and differentiation time constants of 20 ns and 100 ns, respectively. This particular image was taken with the GPM operated with Ne/$CH_4$(5%) at 0.7 bar and ~180 K. The voltage across THGEM1 was 700 V, with 430 V across THGEMs 2 and 3 (gain ~$1\times10^5$); the transfer and induction fields were all 1 kV/cm.

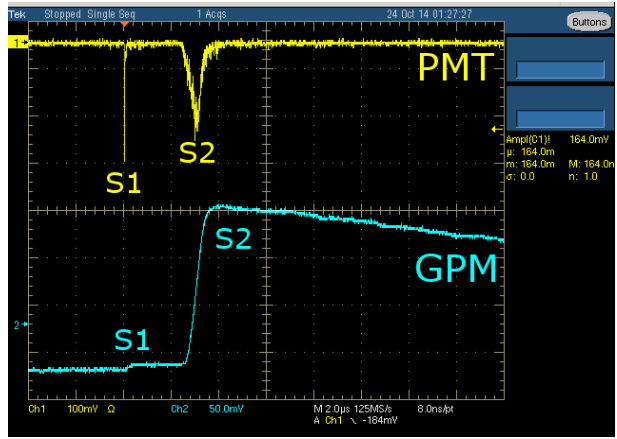

**Figure 4:** S1 and S2 signals from the PMT and GPM (the latter – through a charge sensitive preamplifier). The GPM was operated here with Ne/CH$_4$(10%) at 1.05 bar and ~190 K, at a gain of ~1×10$^5$.

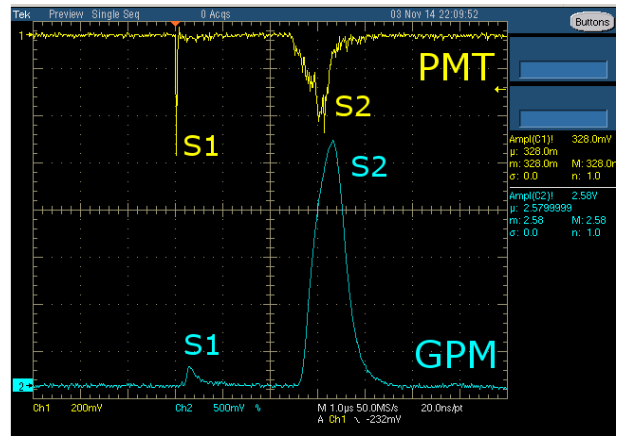

**Figure 5:** S1 and S2 signals from the PMT and the GPM (the latter – through a timing filter amplifier). The GPM was operated here with Ne/CH$_4$(5%) at 0.7 bar and ~180 K, at a gain of ~1×10$^5$.

### 3.2 Gain and stability

A GPM deployed within an array of photosensors in a dual-phase noble-liquid DM detector should have a large dynamic range; it should be capable of consistently recording both single S1 photons with high detection efficiency and S2 signals comprising thousands of photoelectrons with minimum discharges. The GPM gain plays a key role in its detection efficiency: for example, a gain of 1×10$^5$ should permit >90% detection of single-photoelectron signals above noise for front-end electronics with a moderate noise level of ~1 fC (~6,000 e$^-$). For a noise level of ~1000 e$^-$, a gain of ~3×10$^4$ would permit ~95% detection efficiency at the level of a 3σ cut, as shown, for example, for a quintuple-GEM prototype in [36].

Gain measurements were performed by shining a D$_2$ UV lamp through a fused silica window near the top of the GPM port. The lamp provided single UV photons at a rate of a few hundred Hz that reached the CsI photocathode by reflection. The gain was estimated by fitting an exponential function to the pulse height distribution of these photons, as shown in figure 6. The position of the GPM S1 peak due to alpha particle scintillation was also recorded on the multichannel analyzer (MCA), serving as a complementary handle to find relative changes in the gain; in particular, this allowed estimating the gain at low GPM voltages, where the exponential fit was no longer possible because of noise limitations in the present setup. During gain measurements the TPC voltages were set to zero, thus preventing the formation of S2 and leaving only alpha-induced S1 signals at a rate of 40 Hz. The typical amplitude of the S1 pulses was much larger than that of the single-photon signals, and thus did not affect the single-photoelectron pulse-height distribution.

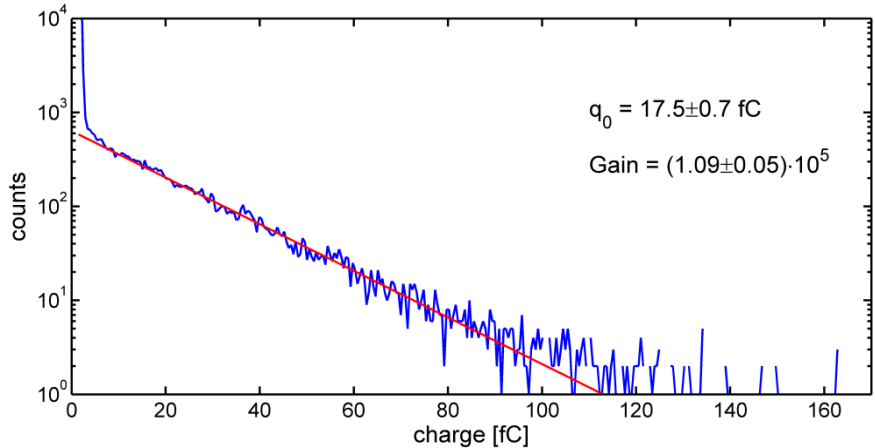

**Figure 6:** Typical GPM single-photon pulse height distribution, fitted by an exponential. The GPM was operated here with Ne/CH$_4$(20%) at a gain of $(1.09\pm0.05)\times10^5$, as deduced from the fit ($q_0$ is the average value of the exponential pulse height distribution; the quoted uncertainty results from varying the fit region). The electronic noise was ~3 fC.

The detector gain (figure 7) depended on the gas composition, with lower voltages required for smaller admixtures of methane for a given gas-multiplication value. The maximum gain obtained at 0.7 bar and 180 K was ~$8\times10^5$ for Ne/CH$_4$(5%) and ~$3\times10^5$ for Ne/CH$_4$(20%); increasing the voltages to higher values resulted in occasional discharges. For both gas mixtures, "asymmetric" THGEM polarization (with higher voltage across THGEM1) proved to be more stable, as shown in figure 7a. Thus, for Ne/CH$_4$(5%) we applied 700 V on THGEM1 with 400-495 V on the THGEMs 2 and 3, while for Ne/CH$_4$(20%) the voltage on THGEM1 was set to 1000 V, with 660-820 V on THGEMs 2 and 3. The transfer and induction fields in both cases were all kept at 1 kV/cm. While operating with the TPC voltages turned on, and thus with S2 signals, the maximal stable gain was lower by a factor of ~2-3 for both gas mixtures. With Ne/CH$_4$(5%) at a gain of $1\times10^5$ and alpha particle-induced S2 signals at a rate of 40 Hz, the discharge probability was found to be of the order of $10^{-6}$. We note that in addition to the alpha particle S2 signals (resulting in a few thousand photoelectrons per event on the GPM), there were also ~20-30 cosmic rays per minute crossing the TPC; these deposited charges resulted in S2 signals up to ~100 times larger than those induced by the alpha particles.

Figure 7b shows two gain measurements performed with Ne/CH$_4$(20%) under similar conditions (0.7 bar, ~190 K) over a period of two months. During this entire time interval, the detector operated in a sealed mode, i.e., with no exchange of the gas. The two curves are consistent to within 7-15% over the range of overlapping voltages, with the higher values obtained in the second measurement (for which the onset of occasional discharges occurred at about two-fold lower gain). The results of the two measurements were consistent for both the $D_2$ lamp and alpha-induced S1 signals, indicating that there were no significant changes in either the gas composition or the CsI QE.

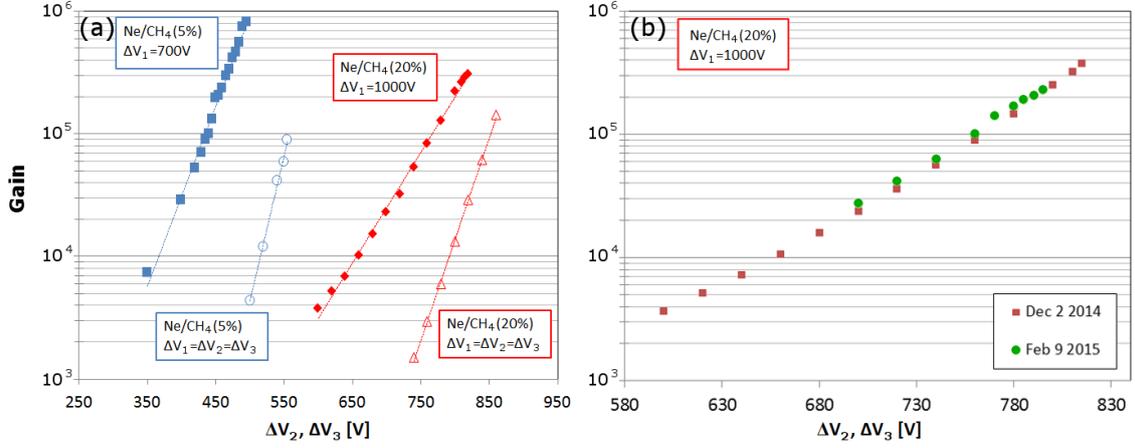

**Figure 7:** (a) Gain curves measured in the triple-THGEM GPM of figure 2, with either equal voltages across all THGEMs or higher (fixed) voltage on THGEM1, in Ne/CH$_4$(5%) and Ne/CH$_4$(20%) at 0.7 bar and 180 K. The gain is plotted against the voltage across THGEMs 2 and 3 ($\Delta V_{2,3}$); in the "asymmetric" cases the voltage across THGEM1 ($\Delta V_1$) was 700 V for Ne/CH$_4$(5%) and 1000 V for Ne/CH$_4$(20%). (b) Gain curves recorded in the sealed mode, separated by two months; Ne/CH$_4$(20%), p = 0.7 bar, T ~190 K.

### 3.3 Energy resolution

Figure 8 shows the GPM pulse height distributions of alpha particle-induced S1 and S2 signals. The detector was operated at 180 K with Ne/CH$_4$(5%) at 0.7 bar; the applied voltages were $\Delta V_1 = 700$ V and $\Delta V_{2,3} = 430$ V and the transfer and induction fields were 1 kV/cm, with a resulting gain of $1 \times 10^5$. For S1 signals we derived an RMS resolution $\sigma/E = 10.9\%$ (by fitting a Gaussian to the entire S1 peak). For S2, the asymmetric shape of the spectrum reflects sum events of alpha and 59.5 keV gamma emissions from the source; a Gaussian fit to the left side of the distribution yields: $\sigma/E = 8.7\%$. Note that in the S2 spectrum in figure 8b, one can clearly see the 59.5 keV peak for events in which the alpha particle is emitted into the source holder and the correlated gamma is emitted into the liquid. The ratio of the alpha and gamma S2 peaks (~5.7) is consistent with the different charge yields from their respective tracks in LXe [3]. The S2 resolution recorded here is comparable to that obtained in XENON100 with PMTs ($\sigma/E=10.0\pm1.5\%$) [18], for a similar number of ionization electrons (~8000) entering the gas phase.

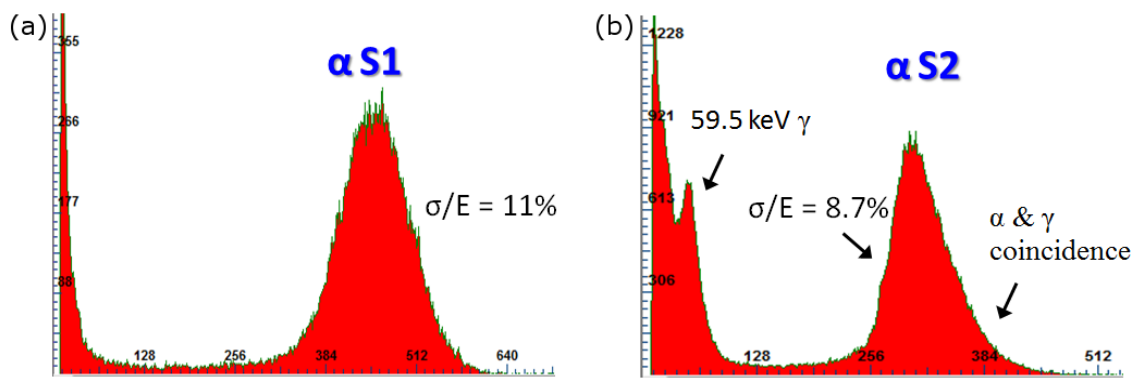

**Figure 8:** S1 (a) and S2 (b) spectra, recorded with the GPM (of figure 2) operated at 180 K with 0.7 bar Ne/CH$_4$(5%) with a gain of 1×10$^5$. The derived RMS resolutions are shown in the figures (see text).

### 3.4 Time resolution

The time delay between the moment of photoelectron emission from the GPM's CsI photocathode (here deduced from the S1 PMT signal) and the signal formation on the GPM's anode depends on the gas composition, pressure and fields (particularly the transfer and induction fields). For Ne/CH$_4$(5%) at 0.7 bar and 180 K with transfer and induction fields of 1 kV/cm, this time difference was ~220 ns, as shown in figure 9; for Ne/CH$_4$(20%) under the same conditions, the time delay was considerably shorter, ~135 ns. This is expected, based on the known increase of the electron drift velocity with the percentage of methane in Ne/CH$_4$ mixtures [46].

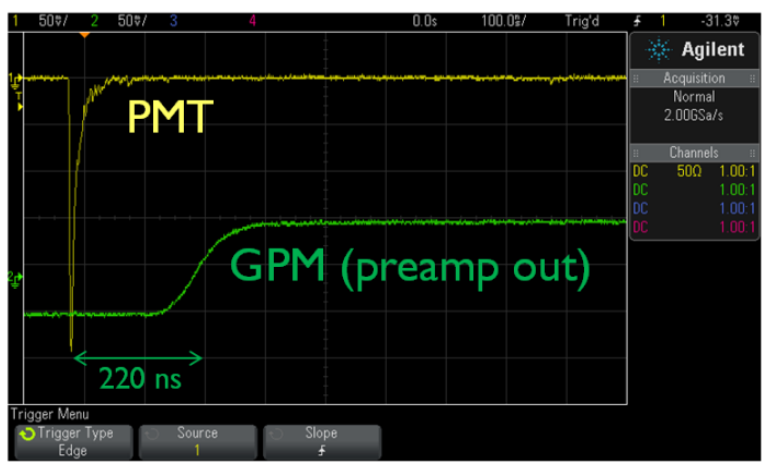

**Figure 9:** The GPM S1 signal from the charge sensitive preamplifier and the corresponding signal on the PMT; the GPM was operated with Ne/CH$_4$(5%) at 0.7 bar and 180 K, with 1 kV/cm transfer and induction fields.

While the GPM response is delayed compared to that of a PMT, its time resolution was shown to be on the nanosecond scale. The GPM time resolution was measured with reference to the PMT signal. The S1 charge signals of the GPM were processed with a timing filter amplifier. Both signals (GPM and PMT) were further processed by a constant fraction discriminator and fed into a time-to-amplitude converter (TAC) – providing a signal whose amplitude is proportional to the time difference between them; the TAC output signals were processed by an MCA (Amptek 8000A). Figure 10a shows the time difference distribution in Ne/CH$_4$(5%), measured at a gain of $3\times10^5$. Figure 10b shows the time resolution (standard deviation of the time difference distribution) as a function of the GPM gain for Ne/CH$_4$(5%) and Ne/CH$_4$(20%) at 0.7 bar and ~190 K. The voltage on the THGEM1 was 700 V for Ne/CH$_4$(5%) and 1000 V for Ne/CH$_4$(20%), and the transfer and induction fields were all 0.5 kV/cm and 1.5 kV/cm for the former, and 1 kV/cm for the latter (the gain was varied by changing the voltages on THGEMs 2 and 3). In all cases, the time jitter decreased with increasing detector gain (i.e., with increasing $\Delta V_{2,3}$), approaching a plateau of ~1.2-1.3 ns RMS. While a full explanation for this behavior requires a detailed simulation study, it may result from improved focusing of the electrons into the holes of THGEMs 2 and 3. For Ne/CH$_4$(20%) the faster stabilization of the time jitter to 1.3 ns (as a function of the overall gain) may be attributed to the smaller diffusion coefficients [46] and higher THGEM fields for the same gain compared to Ne/CH$_4$(5%). We emphasize that these measurements were performed with alpha particle-induced S1 signals resulting in ~170-200 photoelectrons per event; the ultimate time resolution should obviously be defined for single-photoelectron pulses. These are expected to be governed by signal-to-noise issues, by the hole-geometry of the THGEM electrodes, and by the electric field at the photocathode surface.

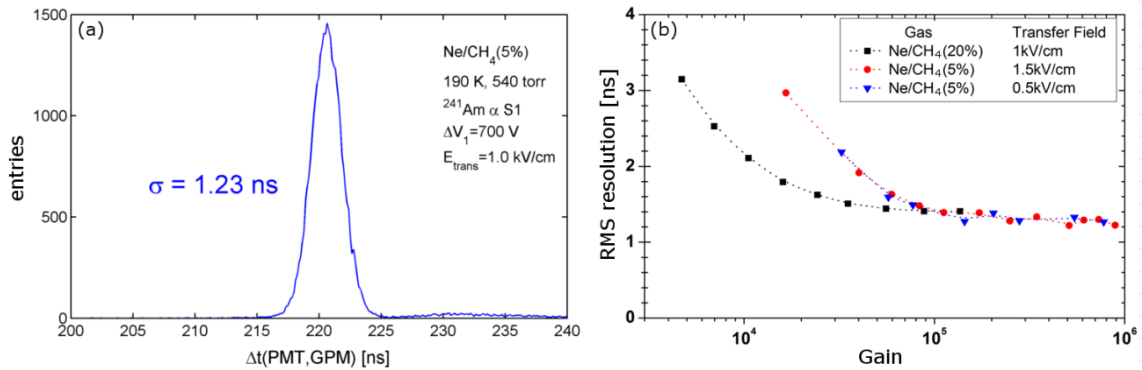

**Figure 10:** (a) Distribution of the time differences between alpha particle-induced S1 signals in LXe measured by the GPM and PMT; σ is the standard deviation. (b) Dependence of σ on the overall detector gain for Ne/CH$_4$(5%) and Ne/CH$_4$(20%) at 0.7 bar and ~190 K.

4. **Photon detection efficiency estimates**

Although the GPM photon detection efficiency (PDE) was not measured directly in this study, it can be estimated for the present configuration by considering the THGEM-electrode geometry and the known (or reference) values of the relevant parameters.

Existing data on the CsI QE and photoelectron extraction efficiency were generally measured at room temperature. That room temperature values are valid also at LXe temperatures was verified by preliminary measurements performed using a simple cryogenic setup, previously described in [39]. The setup comprised a vacuum chamber immersed in ethanol mixed with LN$_2$ at ~170-180 K. The sample was a 300 nm-thick CsI photocathode deposited on Al-coated polished stainless steel, illuminated by a Hg(Ar) lamp (with the main contribution at 185 nm); the photocurrent was measured both in vacuum and under flowing

Ne/CH$_4$(5%). Figure 11a shows good agreement between the photocurrents measured in vacuum at room temperature and at ~175K. Figure 11b shows the extraction efficiency [31] for Ne/CH$_4$(5%), measured under three different conditions: (1) at 295 K and 1.07 bar; (2) at 176 K and 1.07 bar; (3) at 295 K and 1.77 bar (where the gas density is the same as in (2)). As in [31] and [43], the extraction efficiency was defined as the ratio between the photocurrents measured in gas and in vacuum at the same field. The extraction efficiency is plotted against the reduced field, $E/n$, (i.e., the field at the CsI surface divided by the number density of the gas at the given temperature and pressure). The results indicate that the extraction efficiency is completely determined by the reduced electric field, and that the only observable result of cooling to LXe temperature is the increase in gas density (by a factor corresponding to the ratio of absolute temperatures).

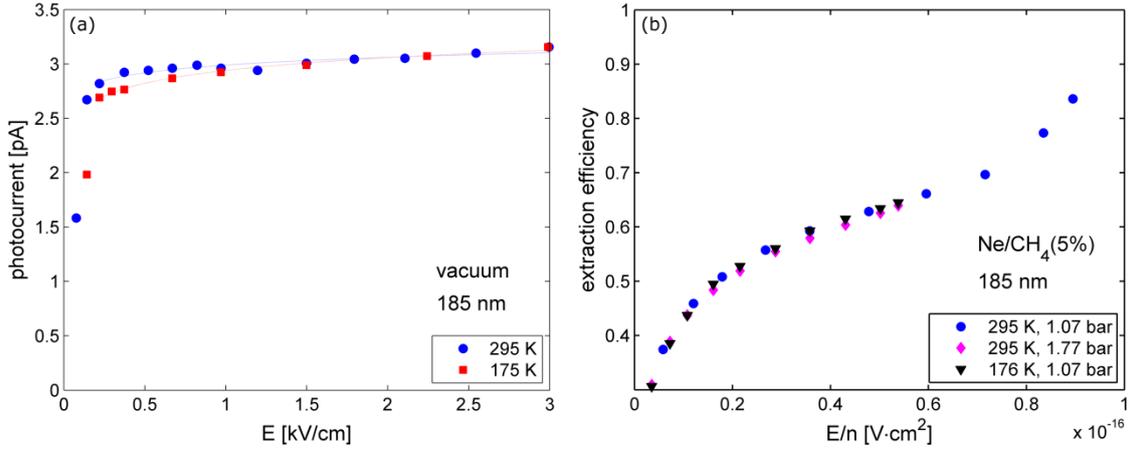

**Figure 11:** (a) Photocurrent of a CsI photocathode illuminated with 185 nm photons in vacuum at room temperature and at 175 K, vs. the electric field. (b) Photoelectron extraction efficiency into Ne/CH$_4$(5%) at room temperature and 1.07 bar, at 176 K and 1.07 bar and at room temperature and 1.77 bar (same gas density as at 176 K) vs. the reduced electric field.

The PDE of a windowed GPM placed in the gas phase of a dual-phase TPC (see figure 2) is determined by the effective quantum efficiency ($QE_{eff}$) defined below, the transmission of photons through the window ($t_W$) and through the mesh ($t_M$), and the probability that the signal is above the electronic noise $f_{SN}$:

$$PDE = t_W \cdot t_M \cdot QE_{eff} \cdot f_{SN} \qquad (2)$$

The effective quantum efficiency, in turn, can be defined as the following product [31]:

$$QE_{eff} = QE \cdot f_{CsI} \cdot \varepsilon_{overall} \qquad (3)$$

QE is the intrinsic quantum efficiency of CsI (here at 175 nm), namely the probability of photoelectron emission from the photocathode into vacuum per impinging photon. Since the QE reaches saturation at fields of a few hundred V/cm (figure 11A), it can be assumed to be constant over the entire CsI area for the THGEM voltages considered here (measurements of CsI QE in previous works were typically done at fields lower than 1 kV/cm [22]). The reference value of the CERN-RD26 collaboration for CsI QE is ~25% at 175 nm [22]. $f_{CsI}$ is the fraction of area covered by CsI, including the hole rims; in the present configuration $f_{CsI} = 0.77$. Lastly, $\varepsilon_{overall}$ is the combined extraction and collection efficiency of transferring the emitted photoelectron from CsI into the THGEM hole in the working gas, averaged over the entire CsI area. Loss of emitted photoelectrons can occur by two processes: elastic backscattering (within ~1-2 mean free paths of the CsI surface) and transverse diffusion along the electron drift trajectory. Previous measurements [31] for a THGEM with 0.3 mm diameter holes, 0.7 mm pitch and 0.4 mm thick FR4, have demonstrated full collection of photoelectrons escaping backscattering for Ne/CH$_4$ mixtures containing 5%-23% methane for THGEM voltages lower than those applied here. Assuming that for larger holes (0.4 mm diameter) and higher THGEM voltages the same holds here, $\varepsilon_{overall}$ should be equal to the overall extraction efficiency, i.e., the probability that a photoelectron emitted from CsI does not backscatter to it, averaged over the entire CsI area:

$$\varepsilon_{overall} = \frac{1}{A_{CsI}} \iint \varepsilon_{ext}\big(E_z(x,y)\big)dxdy \qquad (4)$$

where $A_{CsI}$ is the CsI-covered area (including the hole rims) over which the integral is performed and $\varepsilon_{ext}\big(E_z(x,y)\big)$ is the local extraction efficiency [31, 43], which depends (for a given gas composition and pressure) on the local magnitude of the electric field.

To find $\varepsilon_{overall}$ employing eq. (4), we used the measured values of the extraction efficiency in Ne/CH$_4$ mixtures at room temperature and 1 bar [43], together with a COMSOL calculation of the field across the THGEM top surface (since in this study we worked mainly at 0.7 bar and 180-190 K, this corresponds to roughly the same gas density as at room temperature at 1.1 bar). Figure 12a shows the relevant extraction efficiency data from [43]. Since the field close to the rims can be considerably higher than 2.5 kV/cm (which is the highest value investigated in [43]), the experimental data were fitted by a logarithmic function of the form: $f(E) = a + b \cdot \log(cE)$ and extrapolated to higher field values (different values of the fitting parameters were used for fields below and above 0.5 kV/cm to achieve an optimal fit). For Ne/CH$_4$(5%), where the measured value of the extraction efficiency starts rising more steeply at ~1.5 kV/cm due to gas amplification [43], the extraction efficiency was conservatively taken as that of the logarithmic fit. When calculating the integral in eq. (4), the maximum allowed value for the local extraction efficiency in the extrapolation region (for $E_z > 2.5$ kV/cm) was 1; limiting it to the extraction efficiency measured at 2.5 kV/cm reduced the overall extraction efficiency by a few percent (e.g., by 2% for Ne/CH$_4$(20%) at $\Delta V_{THGEM} = 1$ kV). The calculated results of the overall extraction efficiency are shown in figure 12b as a function of the THGEM voltage. The field was calculated under the assumption that the CsI-covered rims are at the same potential as the Cu layer. The calculation was done for two Cu-clad thicknesses: 64 μm (solid curves), as was used in the present study, and 10 μm (dotted curves) – which we plan to use in future investigations. The THGEM geometry was the one investigated here, namely: $t = 0.4$ mm, $d = 0.4$ mm, $a = 0.8$ mm and $h = 50$ μm. The rim edges were modeled as conical surfaces at 45°, with an inner diameter of $d+2h$. The circular marks on the overall extraction efficiency curves are the values corresponding to the experiments performed with Ne/CH$_4$(5%) with $\Delta V_{THGEM} = 700$ V ($\varepsilon_{overall} = 0.63$) and Ne/CH$_4$(20%) with $\Delta V_{THGEM} = 1000$ V ($\varepsilon_{overall} = 0.77$).

With the calculated overall extraction efficiencies, the nominal effective QEs of the present configuration (assuming QE=25% at 175 nm) for the voltages and pressure applied here, should be 12.1% for Ne/CH$_4$(5%) and 14.8% for Ne/CH$_4$(20%). Considering the transmission of the window (90%) and mesh (85%), the PDE for normally-incident photons, assuming $f_{SN} = 0.95$ for sufficiently high gain and low-noise electronics, would be 8.8% for Ne/CH$_4$(5%) and 10.8% for Ne/CH$_4$(20%) (this can be further improved as discussed below).

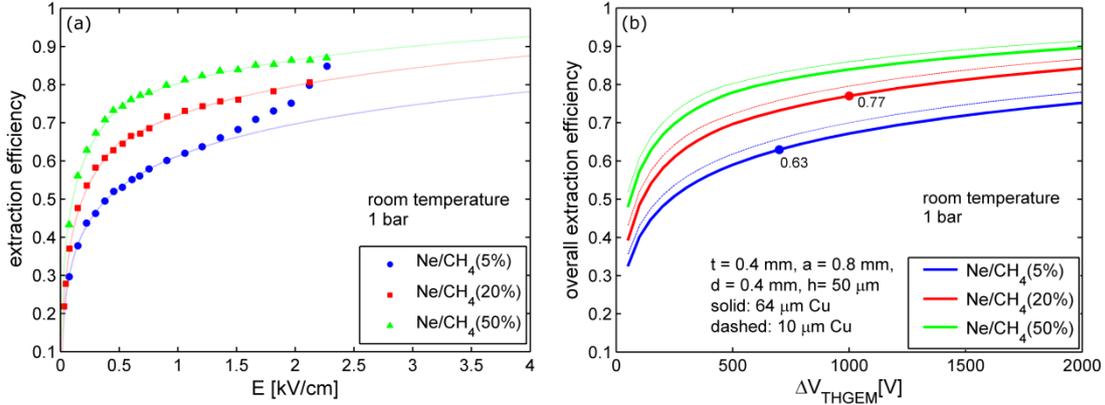

**Figure 12:** (a) Extraction efficiency of CsI into Ne/CH$_4$ mixtures at room temperature and 1 bar; symbols are measured data from [43] and the curves are fitted logarithmic functions used for extrapolation to higher fields (without considering gas amplification in Ne/CH$_4$(5%)). (b) Overall extraction efficiency (eq. (4)) for the THGEM geometry used in the present study. Circles represent the current working voltages with the corresponding efficiencies. Solid lines are for 64 μm-thick Cu clad on the THGEM electrodes (as in the present study) and dotted lines are for 10 μm-thick Cu.

## 5. Summary and discussion

In this work we have demonstrated the detection of both primary and secondary scintillation signals from a dual-phase LXe TPC with a large-area cryogenic gaseous photomultiplier. The detector configuration studied here, with a UV-transparent window and a reflective CsI photocathode deposited on the first amplification stage of a triple-THGEM structure, could potentially be suitable for photosensors located above the xenon vapor phase in such TPCs. A key observation in this respect was the GPM's ability to stably record signals over a very broad dynamic range: at a gain of ~$10^5$ the detector recorded both single photons and large S2 signals comprising thousands of photoelectrons, with a discharge probability of the order of $10^{-6}$. The GPM's energy resolution for alpha particle S2 signals (~9% RMS) was shown to be equivalent to that of the XENON100 dual-phase detector equipped with PMTs, for the same number of ionization electrons (~8000). The RMS time resolution, derived from S1 signals, was demonstrated to be on the nanosecond scale for ~200 photoelectron signals.

As discussed above, the estimated nominal photon detection efficiency (PDE) of the GPM configuration studied here, assuming 25% QE of CsI (in vacuum) at 175 nm and 95% probability of generating a signal above noise, is 10.8% in Ne/CH$_4$(20%), for normally incident photons. This value may be increased by several means. One straightforward improvement would be replacing the present 85% transparent mesh electrode by one of higher optical transparency (98% should be feasible); in addition, THGEM electrodes of similar geometry but with a 10 μm-thick Cu layer on the THGEM1, would enhance the photoelectron extraction efficiency, as seen in figure 12b. Applying these simple modifications and operating with Ne/CH$_4$(20%) at 0.7 bar with 1 kV applied across THGEM1 should increase the PDE to ~13%. An additional improvement would result from an increase in the percentage of methane in the gas mixture to ~50%. Based on figure 12, applying, for example, 1.6 kV across THGEM1 with Ne/CH$_4$(50%) would allow reaching an overall extraction efficiency of 90%, pushing the PDE further to ~15%. Note that the higher methane content would likely require the addition of a fourth amplification stage to maintain sufficiently high gain at reasonable applied voltages per stage. Further enhancement of the PDE could be achieved by modifying the geometry of THGEM1; for example, the use of electrodes with 0.3 mm diameter holes with 0.7 mm or 0.8 mm pitch, would increase the fraction of area covered by CsI from the present 77% to 83% and 87%, respectively. Electric field calculations show, however, that this would come at a price of a somewhat lower field on the THGEM surface, requiring careful optimization of the detector parameters. Lastly, a reduction in the GPM gas pressure could have an additional positive effect on the PDE, as the extraction efficiency is determined by the reduced field.

The overall PDE of a GPM array would be determined by the product of the PDE of an individual detector module and the array fill factor (ratio between the photosensitive area and total instrumented area). The latter would be set by the detector's geometry and size which, in turn, would be affected by mechanical considerations (involving, in particular, the pressure difference across the window). For example, modest-size square 150×150 mm$^2$ GPM units with a 5 mm wide dead region near the detector wall would provide an array fill factor of ~87%; thus, based on the above considerations, overall PDE values of ~13% could be reached. Considering the recent demonstration of a high probability (18-24%) for double-photoelectron emission in PMTs with Bialkali photocathodes at 175 nm [47], the overall PDE of such a GPM array would be similar to that of the top PMT array in LUX [19, 48] and XENON1T [49-52] (both with a fill factor of ~50% and overall PDE of ~12.5%); it will be roughly twice larger than in XENON100 [8] (with a fill factor of 44% and overall PDE of ~6% at 175 nm).

Using pixilated GPMs as the top array of photosensors of a dual-phase TPC could lead to a major improvement in the position reconstruction accuracy in the *xy* plane. As recently demonstrated in [53] for a small LXe TPC, sub-mm *xy* resolution can be achieved with closely packed 1" square pixels (PMTs) for 1.3 MeV gammas. Similar accuracy – for lower energy events – will be readily achievable with pixilated GPMs; the choice of pixel size is arbitrary and will eventually be limited by practical considerations regarding the readout electronics. For comparison, the closely packed array of 2" PMTs in LUX provides an RMS position resolution of ~3-7 mm [54] depending on the depth and energy of the interaction (the former affects the degree of transverse diffusion of the electron cloud). The position resolution in XENON1T with 3" PMTs is expected to be somewhat worse. While a spatial resolution of ~1 mm is not strictly required for gross fiducialization of the LXe target, it may prove to be useful for applying local corrections depending on the exact *xy* position of the event, with possible improvement in the experiment's calibration techniques. In particular, this may be important in future few-meter diameter TPCs, where some internal support structures for the gate and anode grids may be required; in such configurations, local corrections for events occurring close to the supporting frame may be useful for maintaining high S2 energy resolution.

In addition to their possible deployment as top sensor arrays of dual-phase TPCs, one could conceive TPC configurations with GPM readout in 4π geometry, as suggested also in [37]. While the top array could comprise reflective-CsI GPMs, those immersed within the liquid (at the TPC bottom and around the drift cage) would require a new design. Total internal reflection on the inner surface of the window would limit the transmission of isotropically impinging UV photons into the GPM to ~20%. Thus, immersed GPMs would require a design with a semitransparent photocathode on the inner side of the window. As shown in [55], the QE of a semitransparent 10 nm-thick CsI layer is ~20%. This is normally reduced by a factor of ~3 because of photon absorption in an underlying metal coating, used to prevent up-charging of the highly resistive CsI photocathode [22, 55]. However, for low-rate applications such as dark matter searches, the underlying metal layer can be replaced by a highly transparent grid of thin metal strips deposited on the window. GPMs with semitransparent CsI photocathodes evaporated on such grids are part of current R&D by our group and will be discussed elsewhere, along with a detailed analysis of their potential application in large $4\pi$ LXe TPCs.

Two particular challenges must be overcome to make GPMs a viable solution for future dark matter searches: (1) they must be made of radiopure materials, with radioactivity per unit area comparable to that of PMTs; (2) for pixilated GPMs, large-scale, radiopure cryogenic readout electronics should be developed. The first challenge can be addressed by replacing FR4 with low-radioactivity substrates such as Kapton (Cirlex), PTFE, or PEEK. Since these have intrinsic $^{238}$U, $^{232}$Th and $^{40}$K radioactivity levels on the order of $10^{-5} - 10^{-3}$ mBq/cm$^2$ for ~0.5 mm thick plates (based on screening campaigns such as [56-58]), their total radioactivity should be at most on the mBq scale in a 150×150 mm$^2$ GPM module with cascaded THGEM electrodes. While care must be taken not to introduce external radioactivity during the production process of the electrodes, one should note that Cu-clad PTFE and Kapton sheets have typical radiopurity of the same order of magnitude as the bare materials [58, 59]. Additional GPM structural materials, such as the window and metal case, can also be expected to have low radioactive content (in particular – high quality quartz has similar radiopurity as PTFE [56]). As for the readout electronics, both issues (cryogenic operation and radiopurity) are engineering-wise solvable; however, an alternative solution could be the deployment of remote "hot" electronics, possibly requiring an increase of the GPM gain to compensate for higher noise levels.

While the present study focused on LXe, windowed GPMs with CsI photocathodes may also be used for the direct detection of liquid argon scintillation light at 128 nm with no wavelength shifting. The CsI QE at this wavelength can reach ~60% [22]. Coupled to a MgF$_2$ window (with ~60% transmission) this could result in a PDE of ~20%, using non-condensable gas mixtures such as Ar/N$_2$(2%) which provides a photoelectron extraction efficiency of ~80% at modest electric fields [60] (the GPM gas in this case should be maintained at a sufficiently low pressure to prevent Ar condensation).

As a final note, GPMs are currently developed for recording primary scintillation in single-phase LXe detectors for neutron and gamma imaging [61, 62]. They may also be used in conjunction with Liquid Hole-Multipliers (LHMs), developed in parallel by our group, as potential sensors for large-scale single-phase TPCs [40, 63, 64].

**Acknowledgments**

This work was partly supported by the Minerva Foundation with funding from the German Ministry for Education and Research (Grant No. 710827), the Israel Science Foundation (Grant No. 477/10), the PAZY Foundation (Grant No. 258/14) and by FCT and FEDER under the COMPETE program, through project CERN/FP/123614/2011. We thank Dr. A. Lyashenko and Dr. H. Wang for their kind assistance in calibrating the reference detector used in this work, and Dr. A. Kish for his assistance with PMT bases. We further thank Mr. B. Pasmantirer and Mr. S. Assayag and members of their respective Design Office and Mechanical Workshop at the Weizmann Institute, for their invaluable assistance in the design and manufacture of the experimental setup. Lastly, we thank Dr. R. Budnik for many insightful comments throughout the study. The research has been carried out within the DARWIN Consortium for future dark matter experiments. A. Breskin is the W.P. Reuther Professor of Research in The Peaceful Use of Atomic Energy.